\documentclass[aps,prl,twocolumn,groupedaddress]{revtex4-1}
\usepackage{graphicx}
\usepackage{dcolumn}
\usepackage{bm}

\begin{document}


\title{Excitation of radiationless anapole mode of isotropic dielectric nanoparticles with tightly focused radially polarized beam}
\author{Lei Wei, Zheng Xi,  Nandini Bhattacharya,  H. Paul Urbach}
 \affiliation{Optics Research Group, Delft University of Technology, Lorentzweg 1, 2628CJ,  Delft, The Netherlands}

\author{}
\affiliation{}


\date{\today}

\begin{abstract}
A high index dielectric nano-sphere can be excited and yet remain radiationless. A method to excite the non-radiating anapole mode of a high index isotropic dielectric nanosphere is presented. With tightly focused radially polarized beam illumination, the main-contributing electric dipole mode and magnetic modes can be zero with only a weak electric quadruple contributing to the total scattering. Further, with a standing wave illumination formed by two counter-propagating focused radially polarized beam under $4\pi$ configuration, the ideal radiationless ananpole can be excited. 
\end{abstract}

\pacs{}

\maketitle



Particles and objects which cannot be seen has been a fascinating possibility both in popular imagination and in the scientific community. Besides the practical implications in modern day security applications it is also of great interest in the general scientific study of the universe\cite{ho2013}. There can be different possible types of these objects which are radiation less. These can be objects which do not scatter incident radiation\cite{kerker1975}. There also can be time varying charge or current distributions or atoms which do not radiate\cite{ wolf1973, wolf1986, wolf1993, Gbur2001}. The study of such non-radiative objects has been a part of fundamental physics for a long time\cite{Ehrenfest1910,Schott1933, Goedecke1964}. Studies have aimed to understand the physics of these objects from the perspective of self-oscillations of a charged particle\cite{Bohm1948}, scattering from non-absorbing particles\cite{kerker1975} and inverse scattering problems\cite{wolf1993, Hoenders1997} to mention a few. Possible applications of this physics for designing non-scattering objects are also being investigated\cite{leonhardt2006}. \\

One such case of a non-radiating sources is the anapole. This was first introduced in elementary particle physics\cite{Zeldovich1957} and recently has been of interest in particle physics again\cite{ho2013}. The anapole mode arises when the electrical dipole and toroidal dipole moment form a nontrivial destructive interference\cite{Dubovik1990, Afanasiev1995, Kaelberer2010}. However, it is until recently that the nonradiating anapole mode is experimentally demonstrated in the microwave range\cite{Fedotov2013} and  in the visible wavelength range\cite{anapole2015}. In Mie scattering theory, under certain conditions a specific multipole mode from the vector spherical harmonics decomposition of the scattering field of an illuminated particle can be completely suppressed. For example, the  scattering coefficient of the spherical electric  dipole moment can be zero at certain wavelength for a high index dielectric nanosphere under plane wave illumination\cite{Bohren1998}. However, at the same time a strong and broadband magnetic dipole resonance can be excited in the high index dielectric nanosphere\cite{Kuznetsov2012, Evlyukhin2012}. Therefore the nonradiating anapole mode of a nanosphere is difficult to observe under plane wave illumination due to simultaneous excitation of electric and magnetic modes. That's why the first experimental demonstration of anapole in the visible wavelength range\cite{anapole2015} is done with a specially designed nanodisk, instead of a sphere. Inspite of constructing the nanodisk for the specific anapole condition, the scattering from the particle cannot be completely reduced to zero due to the simultaneous excitation of a magnetic quadrupole mode.   Instead of specifically designed structures\cite{anapole2015, liu2015}, in this work we propose an alternative way to excite the anapole mode of an isotropic nanosphere with tightly focused  radially polarized beams. The focused radially polarized beam, which has a pure electric field $E_z$ and zero magnetic field at the focal point\cite{novotny2012principles} can efficiently excite electric dipole and toroidal moment of the induced current at the anapole conditions while keeping the magnetic modes suppressed. In addition we show that when two focussed radially polarized beams under $4\pi$ configuration illuminate the nanosphere an ideal non-radiating anapole is excited.\\
\begin{figure}[!h]
\centering
{\includegraphics[width=7.6cm]{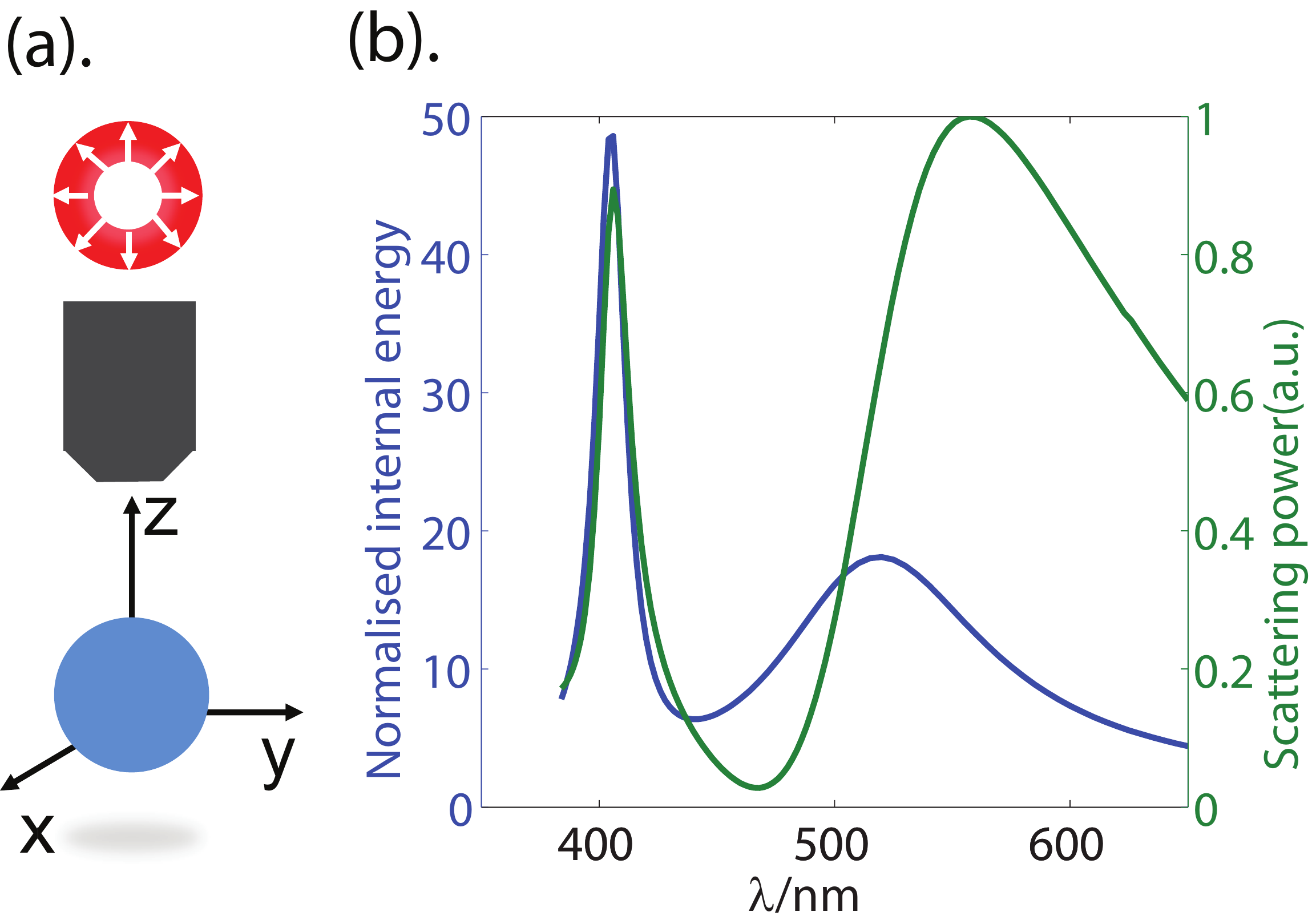}}
\caption{(a). An dielectric nanosphere of refractive index n=3.5 and radius $r_0=100$ nm is illuminated by a focused radially polarized beam in an optical system with numerical aperture NA=0.86; (b). Normalized energy inside the particle and the scattering power for a dielectric nanosphere at the focus of a radially polarized beam. }
\label{fig1}
\end{figure}
The configuration of the proposed anapole excitation is shown in Fig.\ref{fig1}(a), where a tightly focussed radially polarized beam illuminates a high index dielectric nanosphere of radius $r_0=100$ nm and refractive index $\rm{n}=3.5$ placed at the focal point. The surrounding medium is air but the result holds for any high index sphere embedded in a low index, homogeneous and isotropic medium. In the numerical implementation, the focal field field of the radially polarized beam is calculated by the Richard-Wolf diffraction integral\cite{EWolf1958} and then imported to a finite element method simulation\cite{wei2007} to calculate the scattering properties of the nanosphere. In Fig.\ref{fig1}(b) we show the calculated spectral dependence of the scattering power and the internal energy of the particle. Here the electrical energy inside the particle $W_{\mathrm{E}}=\frac{1}{2}\int(\mathbf{E}\mathbf{E}^*)d\mathbf{r}$ is normalized to the focal energy  $W_{\mathrm{f}}=\frac{1}{2}\int(\mathbf{E_{f}}\mathbf{E_{f}}^*)d\mathbf{r}$ within the same volume, when the particle is absent. We can clearly see from the scattering power spectrum in Fig.\ref{fig1}b that only electric modes are excited by the proposed focused radially polarized beam illumination. This selective excitation can be attributed to the special field distribution of the focused radially polarized beam, which has been shown in previous works\cite{Wozniak2015,Das2015, Xi2016}. We also observe from Fig.\ref{fig1}(b), that at $\lambda=464$ nm, the scattering power has a minimum but the internal energy of the particle is 8 times higher than the energy at the focal volume of the illumination, when no particle is present. This is a clear indication that an induced current distribution exists inside the particle, while very little radiation is present outside. This is a clear signature of that an anapole mode is excited.\\
In order to confirm that an anapole moment is excited, the field inside the particles is numerically calculated and Cartesian multipole moment analysis\cite{anapole2015} is applied on the induced current $\mathbf{J}=-\frac{i}{\omega}(n^2-1)\mathbf{E}$ inside the particle. The focused radially polarized beam induces an inhomogenous current distribution inside the particle. It excites not only the Cartesian dipole moment $\mathbf{P}=\frac{i}{\omega}\int \mathbf{J}d\mathbf{r}$, but also the toroidal dipole moment $\mathbf{T}=\frac{1}{10c}\int[(\mathbf{r}\cdot \mathbf{J})\mathbf{r}-2r^2\mathbf{J} ]d\mathbf{r}$. The toroidal dipole moment, featuring an electric poloidal current with a circulating magnetic field, has exactly the same far field scattering pattern as dipole moment. An anapole is excited if the condition $\mathbf{P}=-ik\mathbf{T}$ is fulfilled , where the electrical dipole moment and toroidal moment form destructive interference between each other, resulting in complete cancellation of their far field radiation. \\
\begin{figure}[!h]
\centering
{\includegraphics[width=8.4cm]{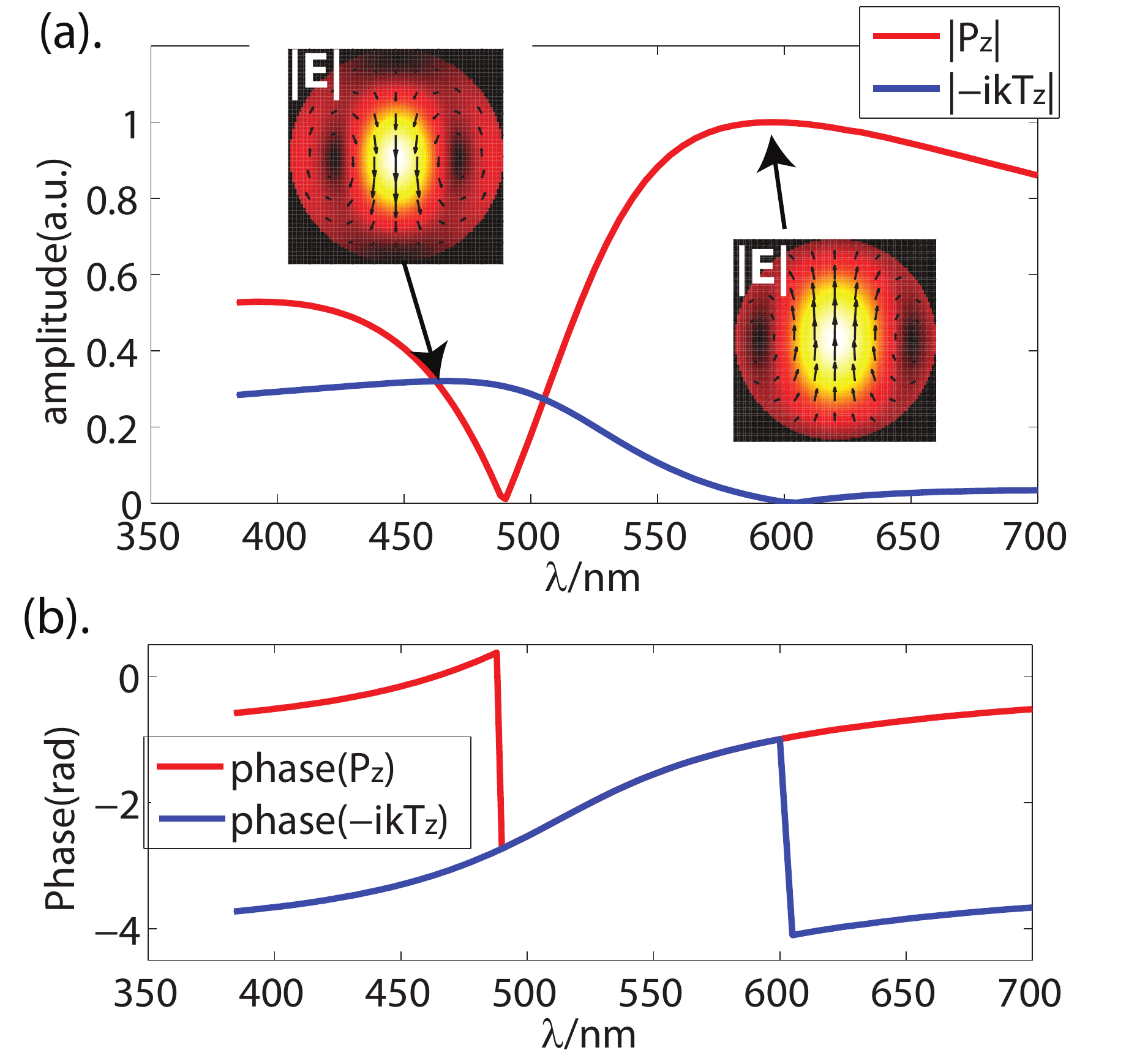}}
\caption{(a) amplitude and (b) phase of Cartesian dipole moment $\mathbf{P}_z$ and toroidal moment $-ik\mathbf{T}_z$ of the induced current inside the particle at focus.  The left inserted figure in (a) shows the induced current distribution at wavelength $\lambda=464$ nm where the anapole mode is excited,  the right inserted figure in (a) shows the pure dipole mode.}
\label{fig2}
\end{figure}
In Fig.\ref{fig2} we illustrate the Cartesian dipole moment and toroidal moment of the induced current inside the particle under illumination of the focused radially polarized beam. Both the electric dipole moment $\mathbf{P}$ and the toroidal moment $\mathbf{T}$ have only nonzero $z$ component, since the focal field of a radially polarized beam in a high NA optical system is dominated by a longitudinally polarized electrical field $E_z$ at the focus and the transverse focal fields are radially polarized. At $\lambda=464$ nm where the anapole mode condition is fulfilled, the scattering fields of dipole and toroidal moment have the same strength but out of phase, resulting in complete cancellation of the far field dipole mode. This proves that indeed an anapole mode isexcited with the focused radially polarized beam.\\
\begin{figure}[!h]
\centering
{\includegraphics[width=8.4cm]{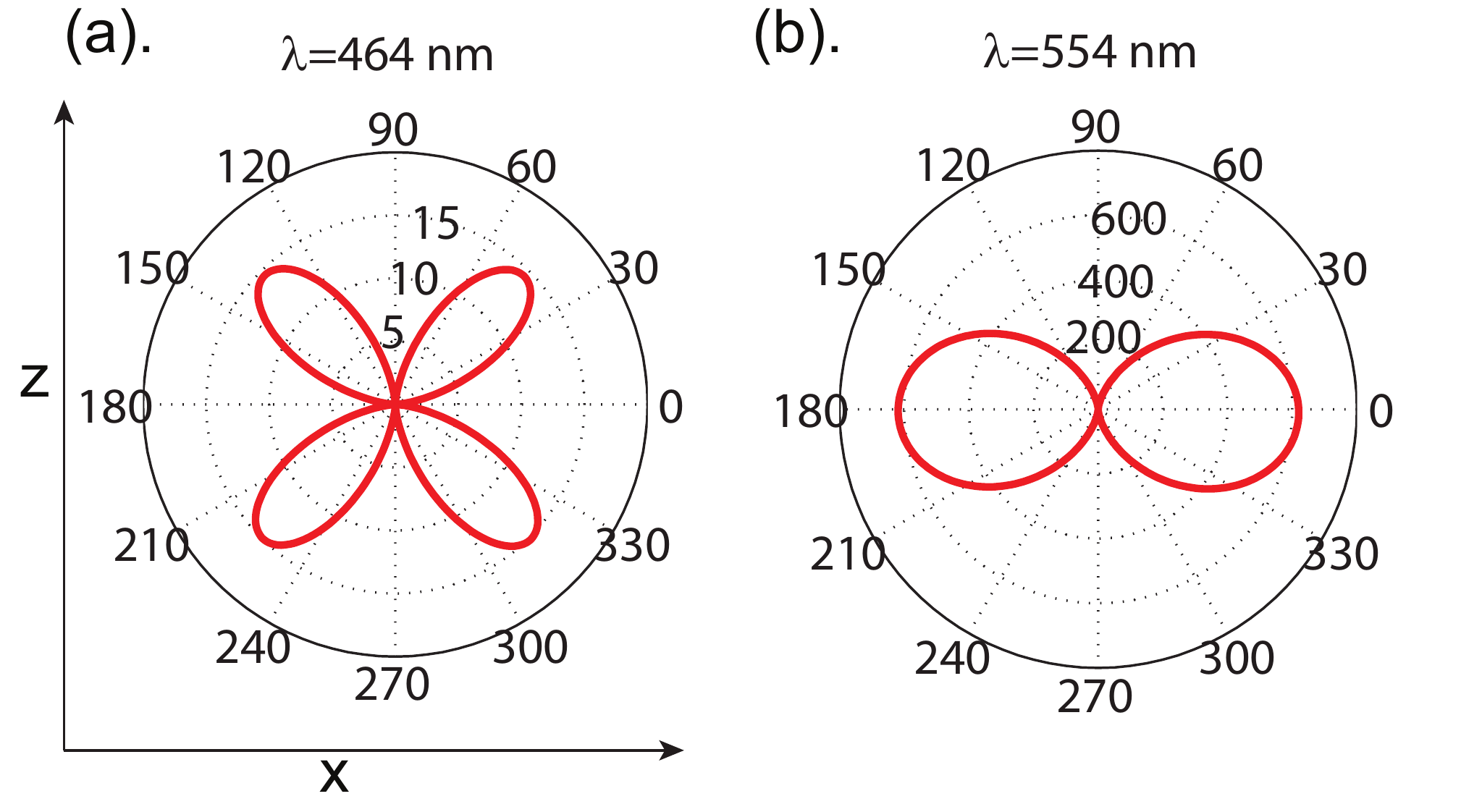}}
\caption{Radiation pattern(radiation power in arbitrary units) of the particle illuminated by a focused radially polarized beam at  the wavelengths of (a) anapole mode and (b) electric dipole mode. 
}
\label{fig3}
\end{figure}
According to the reciprocity theorem, an ideal non-radiating source can not be excited externally by a propagating wave\cite{wolf1993}. As the single focused radially polarized beam has a net energy flow along the z direction, the scattering of the particle can not be completely zero. This can be seen in Fig.\ref{fig3}: although the peak scattering power at $\lambda=464$ nm is 40 times weaker compared to the one of electric dipole resonance at $\lambda=554$ nm, it is not zero, due to the influence of a weak electric quadruple radiation. Since both electrical dipole and toroidal moment of the induced current have only z component, one can detect strong scattering in the z=0 plane at the spherical electric dipole resonance $\lambda=554$ nm. However, no scattering can be detected in this plane when the anapole condition is fulfilled at $\lambda=464$ nm, as shown in Fig.\ref{fig3}(a). \\
\begin{figure}[!h]
\centering
{\includegraphics[width=8.4cm]{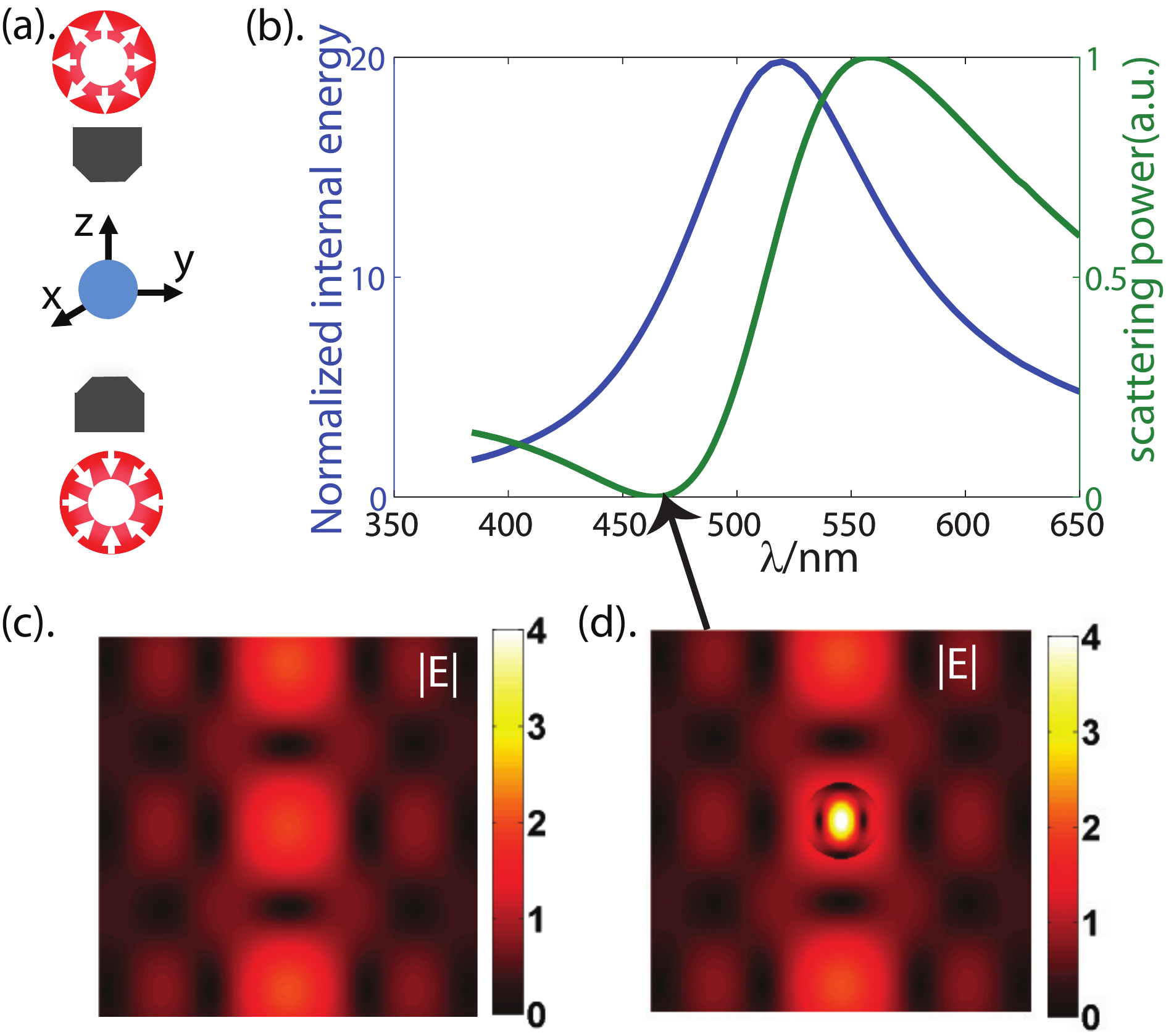}}
\caption{(a). A high-index dielectric nanoparticle of refractive index $n=3.5$ and radius $r_0=100$ nm is excited under $4\pi$ configuration, where the two counter-propagating radially polarized beam have exactly the same intensity but $\pi$ phase difference; (b). The normalized internal energy and the scattering power of the particle excited under the $4\pi$ configuration; (c). The focal electricl field form by the two counter-propagating radially polarized beam under $4\pi$ configuration; (d). Total electric field of the particle excited by the focal field in (c) at the radiating anapole condition $\lambda=464$ nm.}
\label{fig4}
\end{figure}
However, an ideal anapole can be realized with a non-propagating excitation field. Such an excitation field can be formed by the interference of two conterpropagating focused radially polarized beams under a $4\pi$ configuration, as illustrated in Fig.\ref{fig4}(a). The scattering power and internal energy of the particle under such illumination is shown in Fig.\ref{fig4}(b). Due to the interference of the two focal fields in the overlapping focal region, a standing wave is formed as Fig.\ref{fig4}(c). If the top beam and the bottom beam have exactly the same radially polarized pupil field except a $\pi$ phase difference, the transverse field of the two focii interfere destructively at the focal plane, but the longitudinal electric field $E_{z}$ are in phase. This leads to a pure spherical electric dipole moment excitation\cite{Das2015} with no contribution of electric quadruple moment compared to the case of the single focused radially polarized beam as can be seen in Fig.\ref{fig1}(b).  For the the case of the two counter propagating radially polarized beams at $\lambda=464$nm, the scattering power of the particle is exactly zero with no perturbation on the focal field outside the particle as can be clearly seen in Fig.\ref{fig4}(d). However, the internal energy is still 8 times higher than the focal energy in the same region when the particle is absent, implying that an induced current is excited but does not radiate at all, i.e. an ideal radiationless anapole mode is excited with this configuration.\\ 
In summary, we demonstrate that it is possible to excite the anapole mode of an isotropic high index dielectric nanosphere by illumination with tightly focused radially polarized beams. With a single focused beam excitation, the anapole dipole mode can be excited but there is still a weak contribution of electric quadruple radiation in the total scattering power. However, it can be suppressed can an ideal radiationless anapole can be excited with a $4\pi$ configuration, where the two conter-propagating radially polarized beams have the same amplitude but a $\pi$ phase difference. This approach shows for the first time how a radiationless anapole mode can be realized inside the un-engineered isotropic spherical Mie particle. This not only gives us a deeper understanding the physics of the non-scattering objects and nonradiating sources but may also enables applications like non-invasive sensing, suppression of spurious scattering, design of invisible object and optical switches where the scattering can be swtiched on and off with the anapole mode.

\end{document}